\pdfoutput=1
\documentclass{article}

\usepackage[preprint,nonatbib]{neurips_2020}

\usepackage[utf8]{inputenc} 
\usepackage[T1]{fontenc} 
\usepackage{hyperref} 
\usepackage{url} 
\usepackage{booktabs} 
\usepackage{amsfonts} 
\usepackage{nicefrac} 
\usepackage{microtype} 
\usepackage{graphicx}
\usepackage{algorithm}
\usepackage{amsmath}
\usepackage{amsthm}
\usepackage{amssymb}
\usepackage{algpseudocode}
\usepackage{xcolor}
\usepackage{subfigure}
\usepackage{float}
\usepackage{hyperref}
\usepackage{tcolorbox}

\usepackage{tikz} 

\usetikzlibrary{arrows}
\usetikzlibrary{shapes}

\tikzset{
    *|/.style={
        to path={
            (perpendicular cs: horizontal line through={(\tikztostart)},
                                 vertical line through={(\tikztotarget)})
            -- (\tikztotarget) \tikztonodes
        }
    }
}

\tikzset{every picture/.style={line width=1pt}}

\definecolor{columbiablue}{rgb}{0.61, 0.87, 1.0}
\definecolor{classicrose}{rgb}{0.98, 0.8, 0.91}
\definecolor{lightgray}{rgb}{0.83, 0.83, 0.83} 
\definecolor{inchworm}{rgb}{0.7, 0.93, 0.36}
\definecolor{backcolor}{rgb}{0.95,0.95,0.92}
\definecolor{antiquefuchsia}{rgb}{0.57, 0.36, 0.51}
\definecolor{antiquebrass}{rgb}{0.8, 0.58, 0.46}
    
\def \E{{\rm E}}

\theoremstyle{plain}

\theoremstyle{definition}

\newcommand{\expnumber}[2]{{#1}\mathrm{e}{#2}}
\DeclareSymbolFont{symbolsC}{U}{pxsyc}{m}{n}
\DeclareMathSymbol{\coloneqq}{\mathrel}{symbolsC}{"42}
\title {Reinforcement Learning and  Deep Stochastic Optimal Control for Final Quadratic Hedging}

\author{
Bernhard Hientzsch \thanks{All contents and opinions expressed in this document are 
solely those of the author(s) and do not represent the view of Wells Fargo Bank NA.} \\
Corporate Model Risk\\
Wells Fargo\\
}

\begin{document}

\maketitle

\begin{abstract}

We consider two data driven approaches, Reinforcement Learning (RL) and Deep Trajectory-based Stochastic Optimal Control (DTSOC) 
for hedging a European call option without and with transaction cost  according to a 
quadratic hedging P\&L objective at maturity ("variance-optimal hedging" or "final quadratic hedging"). 
We study the performance of the two approaches under various market environments (modeled via the Black-Scholes and/or
the log-normal SABR model) to understand their advantages and limitations. 
Without transaction costs and in the Black-Scholes model, both approaches match the performance of the variance-optimal 
Delta hedge. In the log-normal SABR model without transaction costs, they match the performance of the variance-optimal 
Barlett's Delta hedge. 
Agents trained on Black-Scholes trajectories with matching initial volatility but used on SABR trajectories match the performance of 
Bartlett's Delta hedge in average cost, but show substantially wider variance.
To apply RL approaches to these problems, P\&L at maturity is written as sum of step-wise contributions 
and variants of RL algorithms are implemented and used that minimize expectation of second moments of such sums.  
\end{abstract}
 
\section{Introduction}
Data-driven hedging has attracted wide and deep interest  
as an application of reinforcement learning (RL) and similar approaches to derivatives pricing and  risk management, 
see for instance \cite{deep_Bellman}, \cite{Kolm_Ritter}, or \cite{Risknet}. 
These approaches have advantages, such as (1) that they can work with observed, simulated, or generated data 
for the behavior of the instruments used for hedging and risk management, rather than proceed in model-specific fashion,
(2) that the value and other characteristics of a trading strategy over time can be easily modeled given such data, 
even for realistic models of incomplete markets where trades cause friction and attract transaction costs, and (3)
that hedging or risk management measures of such strategies can be easily computed and used as objectives 
to be optimized over or controlled for. 
These measures could quantify how the strategies perform as measured against book-keeping or market prices of 
the to-be-hedged instruments or quantify the profits and losses when following that strategy to hedge or 
risk-manage the to-be-made or to-be-received payoffs for the to-be-hedged portfolios. In the later case, the behavior of such 
strategies can be modeled and optimized over even as consistent prices or sensitivities might be difficult or
impossible to compute in conventional models with conventional methods. This allows the training of data-driven
agents that optimize said hedging and risk management measures and that perform as well as or better than conventional 
sensitivy-based and model-based hedging approaches. 
Data-driven agents can also take other factors into account and automate or support hedging in situations
where conventional hedging approaches require ad-hoc solutions and tweaks by traders and risk managers.

Several banks and other players in the financial industry have started to investigate and apply these
new approaches to hedging and risk management, such as reported in \cite{RisknetJPM}.  However, the resulting
agents, their behavior, and associated risks of use, including model risk, need to be studied and understood,
so that appropriately designed and understood data-driven hedging and risk-management solutions can be applied 
safely and successfully. Here, we will study how RL and Deep Trajectory-based Stochastic Optimal Control (DTSOC) 
approaches can be applied to hedging problems, in particular hedging to control the quadratic deviation (variance)
of the final profit-and-loss after having made and received all payoffs of the instrument or portfolio to be 
hedged ("quadratic final hedging").

\subsection{Scope of This Work}

This paper concentrates on hedging strategies that control the quadratic deviation (variance)
of the final profit-and-loss after having made and received all payoffs of the instrument or portfolio to be 
hedged ("quadratic final hedging"). The setting will apply to other trading strategies with other objectives also. 

To define the setting, we will model how the prices of the hedging instruments evolve, how the trading strategy 
to do so performs (allowing transaction costs), and define the final objective to be optimized over in terms of 
the prices, costs, etc. involved in the strategy. 

To set up training of agents for final quadratic hedging, one only needs to model the prices of the hedging instruments,
the trading strategy, and the payoffs of the to-be-hedged instrument or portfolio. This is different from the setting 
considered in \cite{fathi2023comparison}, where one also needs to model the prices of the to-be-hedged instruments 
either with a consistent model or a book-keeping model. In situations where the to-be-hedged instrument and closely
related instruments are not liquidly quoted nor traded (either on exchange or on some OTC system) and where no appropriate 
and flexible enough book-keeping models capturing the important features of the instruments and the trading (such as 
transaction costs) are available, final quadratic hedging is applicable (or anything that is defined purely in 
terms of a strategy involving the modeled prices of the hedging instruments) while step-wise mean-variance hedging
would not be applicable. 

Here, we assume we have a model of how the appropriate risk factors and prices for the hedging instruments evolve as 
needed to compute the hedging objectives. This means that we have some implementation that can generate as many trajectories 
of these instrument prices and other risk factors as needed. Based on this data, one can simulate the trading strategy
and determine its performance and thus the objective we will optimize over. 
It is assumed that the trading from the strategies does not impact 
how the risk factors and the prices of the hedging instruments evolve but only how the trading strategy and its results
evolve, as is commonly assumed. We only assume that we can obtain trajectories as needed and do not need any information 
about how the model is obtained, setup, or run.  Thus, either a well-specified traditional model could be used or 
some generative model trained on appropriate data as in  \cite{boursin2022deep} or \cite{Cohen_neural_SDE}.
We will work here with trajectories simulated from a Black-Scholes model or a log-normal SABR model while leaving
synthetic data from generative models to future work. 

We will test and validate our agents on simulated data from such models. Whether agents trained on simulated or generated 
data can perform well on observed data is an interesting research question and is left to future work. There is some 
work \cite{boursin2022deep} that shows that at least in some circumstances agents trained on simulated or 
synthetic data can perform well when run against observed market data.

The agents themselves are operating in a model-free, data-driven setting, since they only are provided 
generated trajectories and rewards rather than details about the model. Given our setting, we are in a
rich data regime (even unlimited data regime), corresponding to a well defined conventional or generative model
that is used to generated as much data as wanted.

\subsection{Related Work}

We discuss here some of the existing work in the literature on the use of Reinforcement Learning and Deep Trajectory 
Based Stochastic Optimal control to pricing and hedging of options. For some broader survey of RL applications in 
finance, see \cite{RLfinance}. 

In \cite{QLBS}, a European call option in the Black-Scholes model (trading at discrete times, no transaction costs) 
is hedged and priced using a $Q$-learning method (there, called QLBS method). 
The objective is minimizing the sum of the initial cash position and the weighted discounted sum of variances of the 
hedging portfolio at all subsequent time steps. This setting and objective 
can be identified with a corresponding MDP setup 
and thus solved with standard RL/DP approaches such as $Q$-learning \cite{QLBS}. 
The price at maturity has to be equal to the final payoff and prices respective $Q$-functions at earlier times
are determined by Dynamic Programming or other standard RL approaches. 
 
\cite{cao2021deep} defines the objective as a combination of the expected final P\&L and the expected variance 
of the final P\&L.  They use versions of $Q$-learning and other methods for the expected sum of contributions and the expected 
square of sums of contributions to address this problem. We will use similar approaches to optimize the variance of 
the final P\&L of the hedged portfolio, but \cite{cao2021deep} did not discuss or apply their approaches to 
final quadratic hedging.  \cite{cao2021deep} test their methods on data simulated from the Black-Scholes and
log-normal SABR model and compare their results to Delta hedging with current implied volatility and Bartlett's Delta
hedging with current implied volatility. 

Many works consider neural networks as a numerical tool to accelerate calculation tasks and to solve generic 
classes of problems, such as model calibration \cite{hernandez2016model}, solving PDEs \cite{Deep_PDE}, 
solving stochastic optimal control problems \cite{han2016deepsc} more efficiently and 
in higher dimensions. 
Among these, we distinguish the line of works initiated in \cite{han2016deepsc} 
where neural networks are used to solve certain stochastic optimal control problems numerically.
More concretely, the DTSOC method in \cite{han2016deepsc} is a setup for trajectory based empirical deep 
stochastic optimal control with both step-wise and final objectives/costs.  
Finally, the "deep hedging" paper \cite{buehler2019deep} presents a trajectory based empirical 
deep stochastic optimal control approach to minimizing some global objectives related to 
replication and/or risk management of some final payoff, according to risk measures such as CVaR/expected 
shortfall, expected utility, etc. While quadratic hedging is mentioned, the method(s) are not 
applied to any variance-optimization/quadratic hedging problem in that paper. 

\section{Framework}

\subsection{General Setup}

We consider here dynamic hedging, trading, and/or risk management problems defined 
in terms of some instrument or portfolio to be hedged or otherwise risk managed. 
In our examples, we assume that the instruments are European options; 
but one could handle portfolios or such or more path-dependent instruments 
in a very similar fashion. 
In these problems, there is an universe of given instruments (including 
cash/bank accounts that allow also negative balances) 
and the agent has to decide how much of each instrument to hold at each time 
to achieve a certain hedging, trading, or risk management objective. 
We assume, as is commonly done, that there is a given fixed set of times at 
which trades can occur and we allow that such trades might incur transaction costs.
We also assume that the strategies are self-financing. The objectives under 
consideration can be defined from contributions at one or several fixed time horizons 
or each trading period and/or trading time might have a contribution associated to it. 
Here, we will consider objectives that are associated with the one final payment time of the to-be-hedged instrument. 
While the objective can be a simple mean-square hedging error as here, it could also be defined based on a 
risk measure (such as CVaR) or some moments or percentiles of an appropriately defined hedging outcome. 

This is a stochastic sequential decision problem. The prices of the hedging 
instruments (here, stocks) will follow some stochastic process with some given model or
generator. These prices can be observed and other appropriate risk factors could 
be computed based on the history of such prices. Now, given these observed prices 
and computed risk factors, the agent needs to decide in time for each trading time 
how much of each hedging instrument to buy and/or sell; or, equivalently, how much 
of each hedging instrument should be held immediately after that trading time; taking 
into account transaction costs and other trade impacts. 
(Thus, holding sizes or trade sizes would be appropriate decision variables.) 
Given to-be-achieved holding size or chosen trade size and the cash at hand before the trade,
the cash remaining after (or being borrowed) can be computed by self-financing and funding strategy, 
the value of the holdings and of the cash at the end of the no-trade interval follows from the stochastic 
evolution of the prices of the hedging instruments and the interest on the cash or loan account, up to final maturity.   

\subsection{Trading Strategy for the Hedging Portfolio} 

We consider the case of a hedging portfolio. This means that we are 
given some description of the to-be-hedged instrument or portfolio. This 
description in general consists of random variates $(P_t)_{t \in \mathcal{T}}$ 
describing the contractually-agreed payoffs under that portfolio,
where $\mathcal{T}$ is the set of their payment times. 
We assume here that we are treating European options: there is only one
payoff at some maturity $T$ which we denote $P_T$. 

As in \cite{fathi2023comparison}, 
it is assumed that we are given "hedging" instruments (say, one or several stocks), 
we denote their prices by $S_t$ (in general a vector), and 
denote the holding of them in our strategy at time $t$ with $H^S_t$ (a vector of corresponding length).
For our examples, we will typically assume a single hedging instrument. 
Finally, we assume that excess cash from the trading strategy is deposited in 
a money-market account with a certain given interest accrual and discount factor 
(and that negative cash balances can be borrowed at the same interest accrual and discount factor). 
We assume that we initially start with some initial cash $Y_0$ but no holdings 
in the hedging instruments (i.e., $H^S_{0^{-}}=0$). 
The trading times are given as $t_i$ with $t_0=0$. If referred to, 
$t_{-1}$ is a time just before time 0 with no holding in hedging instruments. 

Then, at each time $t_i$ when one needs to decide what to trade, 
one can observe the prices $S_{t_i}$ of the hedging instruments. At the same 
time the holdings in these instruments are still given by $H^S_{t_{i-1}}$.  
Now one has to decide how much ($H^S_{t_i}$) shall be held for the 
next no-trading period. Here, we assume that these trades are made instantaneously at $t^+_i$. 
The minimal state $s_i$ for this decision has to include at least 
$(t_i,{S}_{t_i},H^S_{t_{i-1}})$.\footnote{One can also add other features and information 
to the state space that the agent could potentially use to make better decisions or that 
allow the agent to be trained more easily. Also, instead of ${S}_{t_i}$, one can 
use $\tilde{S}_{t_i}$ as defined below, which is ${S}_{t_i}$ discounted back to time 0.} 
The action has to either describe $H^S_{t_i}$ or how to obtain $H^S_{t_i}$ from $H^S_{t_{i-1}}$ .

Consider time $t_{i+1}$, right before the trade: The value of the hedging 
part of the strategy then is \cite{fathi2023comparison} 
\begin{equation}
Y_{t_{i+1}} = \frac{\left( Y_{t_{i}} - H^S_{t_i} S_{t_i} - \mathsf{TC}(H^S_{t_i},H^S_{t_{i-1}},S_{t_i}) \right)} {\mathsf{DF}_{i,i+1}} + H^S_{t_i} S_{t_{i+1}} 
\end{equation} 

This reflects that at the previous trade time $t_i$, the portfolio was rebalanced 
to the appropriate holdings $H^S_{t_i}$, transaction costs were charged 
$\mathsf{TC}(H^S_{t_i},H^S_{t_{i-1}},S_{t_i})$, the thus resulting cash balance 
$Y_{t_{i}} - H^S_{t_i} S_{t_i} - \mathsf{TC}(H^S_{t_i},H^S_{t_{i-1}},S_{t_i})$ 
attracts step-wise interest (interest accrual corresponds to dividing by 
appropriate step-wise discount factor $DF_{i,i+1}$). 
At the same time, the holdings in the hedging instruments are now 
valued based on the new price of the hedging instruments. 

It helps to rewrite the equation as follows as in \cite{fathi2023comparison}:
\begin{equation}
Y_{t_{i+1}} = \frac {1} {\mathsf{DF}_{i,i+1}} Y_{t_i} + H^S_{t_i} \left( S_{t_{i+1}} - \frac {S_{t_i}} {\mathsf{DF}_{i,i+1}} \right)
- \frac{\mathsf{TC}(H^S_{t_i},H^S_{t_{i-1}},S_{t_i})} {\mathsf{DF}_{i,i+1}} 
\end{equation} 

In this completely linear setting without stochastic interest rates 
(and assuming that the transaction cost is linear in the instrument 
prices for positive multipliers $c$,
$ \mathsf{TC}(\cdot,\cdot,cS) = c \mathsf{TC}(\cdot,\cdot,S)$), 
it is possible to "hide" the discounting in the value definitions in the following sense: 
with $\tilde{Y}_{t_i} = \mathsf{DF}_{0,i} Y_{t_i}$ and $\tilde{S}_{t_i} = \mathsf{DF}_{0,i} S_{t_i}$, we have
\begin{equation}
\tilde{Y}_{t_{i+1}} = \tilde{Y}_{t_i} + H^S_{t_i} \left( \tilde{S}_{t_{i+1}} - \tilde{S}_{t_i} \right)
- \mathsf{TC}(H^S_{t_i},H^S_{t_{i-1}},\tilde{S}_{t_i})  
\end{equation} 

In particular, this means for the increment
\begin{equation}
\tilde{dY}_{t_{i+1}} = H^S_{t_i} \left( \tilde{S}_{t_{i+1}} - \tilde{S}_{t_i} \right)
- \mathsf{TC}(H^S_{t_i},H^S_{t_{i-1}},\tilde{S}_{t_i}) 
\end{equation} 
and we thus do not get any contribution or impact of the 
initial cash position $Y_0$ beyond an additive shift.

\subsection{Final Quadratic Hedging Objective}
We will also discount the final payoff and denote it $\tilde{P}_T$.
The discounted final hedging error is now:
\begin{equation}
\tilde{L}_T = \tilde{P}_T - \tilde{Y}_T 
\end{equation}
and we look to minimize the mean square hedging error MSHE ("quadratic hedging"):
\begin{equation}
\mathsf{MSHE}\left(Y_0,H^S_\cdot\right) = \mathbb{E} \left[ \tilde{L}^2_T \right]
\end{equation}
where the expectation is under the given model respective generator 
for the underlying hedging instruments. 
This hedging is also referred to as variance optimal hedging 
(where the variance is of the final hedging error). 
We typically avoid this name except when referring to the literature 
since it does not further specify which variance and use "final quadratic hedging" 
since this makes it clearer that we optimize over the expectation 
of the square of the final hedging mismatch.  

Extensions to stochastic interest rates and discounting, differential rates, 
more involved funding policies are possible, 
require further notation and details, and can be handled similarly, but are 
not necessary for the settings we discuss here. 
Differential rates and more involved funding policies could be included in 
the framework by adding bank accounts for positive 
and negative balances and other funding instruments to the hedging instruments 
and adding constraints to the decision variables 
(only positive holdings of positive bank account, only negative holdings 
of negative bank account i.e. bank debits or loans, 
restricting secured loans/repos to the amount held in the corresponding collateral, etc.).

Under some circumstances, the strategy might spend a lot of transaction 
costs to decrease the final hedging error 
(and thus requiring larger initial wealth $Y_0$). In such cases, one might 
constrain $Y_0$ or also minimize over initial 
wealth, or one might constrain total transaction cost 
- or also minimize over accumulated total transaction costs \cite{fhkpx2023ip}.

Since the impact of initial wealth is only additive and separated, 
we can compute the optimal $Y_0$ for each 
hedging strategy $H^S_\cdot$. The formulation also covers the case
in which $Y_0$ has been chosen from the outside 
(prices set by other mechanism or the instrument is on the run and 
we start with the existing cash holdings (and/or stock holdings) 
assigned to the hedging of our portfolio) where we only 
optimize over the strategy for the remaining time until 
maturity.

\subsection{Modeling the Hedging Instruments}

The formulation of the objective function in this paper only depends 
on the final hedging error - i.e., the difference between the 
contractually agreed (discounted) payoff $\tilde{P}_T$ and the 
final discounted value of the hedging strategy $\tilde{Y}_T$. 
$\tilde{Y}_T$ in turn depends on the discounted prices of the 
hedging instruments and transaction costs earlier, but 
does not require any further information about the models 
for these instruments and prices. 

The agent only  needs to determine how many units to hold in each of 
the hedging instruments after rebalancing at each rebalancing time $t_i$, i.e. $H^S_{t_i}$.
Alternatively, one could specify the units to be sold or bought at 
any given time together with the initial position; or assume that the 
amount sold or bought is proportional to the time between trades or 
some other variable and then specify that proportional trading rate
in that setting. 
The minimal state for modeling this decision problem are these
 discounted prices (hedging instruments), 
the amount of the hedging instruments, time, and whatever state 
the models for these prices and the transaction costs need. 

As for the models for the prices of the hedging instruments, 
one can model them under pricing or observational measure,
with some differences in parameters and calibration or fitting. 
For the hedging instruments, one can use conventional quantitative 
finance models, possibly with hidden factors, such as Black-Scholes, 
Local Volatility Model, SABR model (with stochastic volatility),
Heston model (with stochastic variance), 
quadratic rough Heston model, etc. 
One could use generative models such as GANs or appropriately 
trained neural Stochastic Differential Equations (SDEs) or similar. 
Finally, one could use historically observed data over one instrument 
or a cross section of similar instruments together with appropriate assumptions 
to generate possible future price movements or future prices.\footnote{See also 
\cite{cohen2021blackbox} for a discussion of possible approaches together with associated model risks.}    

For the to-be-hedged instrument, we only need to know the payoff at maturity 
and we do not need any model or method to compute the  price of the to-be-hedged 
instrument at any other time, unlike in hedging and risk management setups that require book-keeping
or reference prices for the to-be-hedged instruments \cite{fathi2023comparison}.
We consider at first simple examples such as the hedging of a short call option that 
was sold to some counterparty or the hedging of a long call option that was bought. 
Here, one only needs to specify the final payoff $P_T = -(S-K)_+$ respectively $P_T = (S-K)_+$.
Black-Scholes models with constant or deterministic volatility and log-normal SABR 
model are considered here, but other models can be easily implemented. 

In formulas, this would mean for the discounted price of the hedging instruments ("stocks") in continuous time
\begin{equation}
d\tilde{S}_t = \sigma_t \tilde{S}_t dW_t 
\end{equation}
with $\sigma_t$ a constant or deterministic time-dependent function (Black-Scholes), or following 
the stochastic volatility SDE of a SABR model
\begin{equation}
d\sigma_t = \frac{\eta}{2} \sigma_t dW^2_t
\end{equation}
and in discrete time, appropriate log-Euler time-discretizations thereof 
(which are actually exact for $\tilde{S}_t$ in this case). 
The Black-Scholes SDE also corresponds to the SDE for the 
underlier in a log-normal SABR model (with $\beta=1$).

If one considers more complicated models based on conventional quantitative finance models, 
one would simulate the (discounted) underlying instruments 
with that model. If these models have additional factors, these factors would be
added to the state. If some of these factors are latent or hidden factors, one would 
need to add some mechanism how these latent 
factors can be estimated or taken into account by some process on observed quantities, 
add these observed quantities to the 
state, and learn agents that only depend on observed and observable quantities, not the latent factors that will in general 
be unknown (and unknowable) to the agent.  
The stochastic volatility in our log-normal SABR test case
can be considered either as a latent, unobserved state or as observed state. 
(We treat it here as a latent state that is not provided as input to
the agent/policy.)

In general, trading in hedging instruments could impact the prices 
in those hedging instruments either temporarily or permanently. 
We assume here that the hedging instruments are traded liquidly and 
that the hedged instrument is such that it can be hedged 
without impacting the prices of the hedging instruments. To a certain 
extent, short-term price impact can be modeled by and 
absorbed into the transaction cost terms.  

As discussed in abstract and introduction, here we focus on 
Black-Scholes and the log-normal SABR stochastic volatility models to investigate the agents 
and algorithms in a setting where the model and the features are 
simple enough, and will consider other models in future work.

\section{Analytical Exact or Approximate Variance-Optimal Hedges}

For some models (Black-Scholes, SABR, Rough Bergomi) and some assumptions (time-continuous trading, 
accurate enough approximations), 
one can derive explicit analytical forms of exact or approximate
variance-optimal hedges that minimize MSHE under those assumptions.
\cite{keller2022bartlett} derives the below formulas (for the case of zero interest rates) and 
applies them to the log-normal SABR and the Rough Bergomi model. 

Assume thus that we have given $C_t$ as the time $t$ price 
for an instrument with payoff $P_T$ (which means in particular 
that $C_T=P_T$) and 
denote the discounted-to-time-zero price as $\tilde{C}_t=D_t C_t$, 
with $D_t$ being the time zero discount factor for time $t$, assuming also
deterministic interest rates.
Similarly denote the discounted-to-time-zero version of $S_t$ by 
$\tilde{S}_t=D_t C_t$. We assume that we simulate and 
operate under a pricing measure $\mathbb{Q}$ under which $\tilde{C}_t$ 
and $\tilde{S}_t$ are square integrable martingales. 
Given any dynamic continuous-time trading strategy $\theta_\cdot$, 
the final discounted value of the hedging portfolio if started from
initial wealth $w$ would be 
\begin{equation}
\tilde{Y}_T = w + \int_0^T \theta_u d\tilde{S}_u 
\end{equation}

The final hedging error is thus 
\begin{equation}
\tilde{L}_T = \tilde{C}_T - \tilde{Y}_T.
\end{equation}
We minimize over the (risk-neutral) mean-square hedging error 
\begin{equation}
\mathsf{MSHE}\left(w,\theta_\cdot\right) = \mathbb{E} \left[ \tilde{L}^2_T \right]
\end{equation}
and the minimizer under the above assumptions is 
called variance-optimal strategy $\theta^{VO}$. 
It is given as the Radon-Nikodym derivative of 
the finite-variation process $\langle\tilde{S},\tilde{C}\rangle_t$ 
with respect to the finite-variation process $\langle\tilde{S},\tilde{S}\rangle_t$:
\begin{equation}
\theta^{VO}_t = \frac {d\langle\tilde{S},\tilde{C}\rangle_t} {d\langle\tilde{S},\tilde{S}\rangle_t}.
\end{equation}

Further assume that $P_T$ represents a call (similar 
derivations work for puts) and that we write its discounted price in terms 
of implied volatility $\Sigma_t$ by the Black-Scholes formula 
for zero interest rates (here denoted by $c_{BS}$)\footnote{
A call payoff discounted back to time zero can be written as 
$D_T(S_T-K)^+ =(\tilde{S}_T-\tilde{K})^+$ 
which can be understood as a call on $\tilde{S}_T$ with 
discounted strike $\tilde{K}=D_T K$, needing no discounting of the payoff .
The Black-Scholes formula then simplifies to 
$c_{\mathsf{BS},t}(\tilde{S}_t,\Sigma_t;T,\tilde{K})=\tilde{S}_t \Phi(d_t^+(\tilde{S}_t,\Sigma_t)) - \tilde{K} \Phi(d_t^-(\tilde{S}_t,\Sigma_t))$. 
Since $\tilde{K}$ and $T$ are fixed once we fix the call option to be hedged, we omit these two arguments to our function since
they are already bound.  }:
\begin{equation}
\tilde{C}_t = c_{\mathsf{BS},t}(\tilde{S}_t,\Sigma_t;T,\tilde{K})=c_{\mathsf{BS},t}(\tilde{S}_t,\Sigma_t).
\end{equation}

This $\Sigma_t$ will be a stochastic process (and as a smooth 
function of martingales, will be a semi-martingale). We introduce
\begin{eqnarray*}
 d_t^{\pm} (\tilde{S}_t,\Sigma_t) &=&\frac{\log(\tilde{S}_t/\tilde{K})}{\Sigma_t\sqrt{T-t}}\pm\frac{\Sigma_t\sqrt{T-t}}{2}\\
 \mathsf{Delta}_t(\tilde{S}_t,\Sigma_t)&=&\Phi(d_t^{+}(\tilde{S}_t,\Sigma_t))\\
 \mathsf{Vega}_t(\tilde{S}_t,\Sigma_t)&=&\tilde{S}_t\phi(d_t^{+}(\tilde{S}_t,\Sigma_t))\sqrt{T-t}.
\end{eqnarray*}

In terms of $\Sigma_t$, we obtain:
\begin{equation}\label{QH_optimal_formula2}
\theta^{VO}_t= \mathsf{Delta}_t(\tilde{S}_t,\Sigma_t)
+\mathsf{Vega}_t(\tilde{S}_t,\Sigma_t)\frac{d\langle\Sigma,\tilde{S}\rangle_t}{d\langle\tilde{S},\tilde{S}\rangle_t}
\end{equation}
with the corresponding Radon-Nikodym derivative in the second term. 

Note that if $\Sigma_t$ is constant or a deterministic function 
of time, the second term is zero and $\mathsf{Delta}_t$ is 
a variance optimal strategy, as in Black-Scholes with constant 
or at most time-dependent volatility $\sigma(t)$.  

Thus, if we have an explicit $\Sigma_t$ and we can compute 
$d\langle\Sigma,\tilde{S}\rangle_t$ and $d\langle\tilde{S},\tilde{S}\rangle_t$ explicitly, 
we have an explicit variance optimal strategy. We can also use 
close approximations $\hat{\Sigma}_t$ of $\Sigma_t$ and would expect the resulting
$\theta^{AVO}_t$ to be a good approximation of $\theta^{VO}_t$.
 
Log-normal SABR for the discounted-to-time-zero stock price $\tilde{S}_t$ is written  
\begin{eqnarray*}
d\tilde{S}_t=&\tilde{S}_t\sigma_t dW_t\\
d\sigma_t=&\frac{\eta}{2}\sigma_t dW^{2}_t,
\end{eqnarray*}
where $d\langle W_t ,W^2_t\rangle_t=\rho dt$ for $\rho\in[-1,1]$, we have 
\begin{eqnarray}
\hat{\Sigma}_t  & = & \sigma_t f(M_t) \\
M_t &=& \frac{\eta}{\sigma_t} \log \left( \frac{\tilde{K}}{\tilde{S}_t} \right)
\end{eqnarray}
with $f$ given in the SABR formula (\cite[equation (15)]{keller2022bartlett} 
and \cite{fukasawa2021rough, fukasawa2022rough,haganetal2002sabr}):
\begin{equation}
f(y) = \frac{y/2}{\log\left(1-\rho\right) - \log\left(\sqrt{1+\rho y+y^2/4} - \rho - y/2\right)}
\end{equation}

The (approximated) Radon-Nikodym derivative can be computed as follows \cite{keller2022bartlett}:
\begin{equation}
\frac{d\langle\hat{\Sigma},\tilde{S}\rangle_t}{d\langle\tilde{S},\tilde{S}\rangle_t} = \frac{\eta}{2\tilde{S}_t} \left( \rho f(M_t) - (\rho M_t+2)f'(M_t) \right)
\end{equation}

Thus, we obtain an approximately variance optimal strategy:
\begin{equation}
\theta^{AVO}_t= \mathsf{Delta}_t(\tilde{S}_t,\hat{\Sigma}_t)+\frac{\eta}{2} \phi(d_t^{+}(\tilde{S}_t,\hat{\Sigma}_t))\sqrt{T-t}  
\left( \rho f(M_t) - (\rho M_t+2)f'(M_t) \right)
\end{equation}
This turns out to be exactly Barlett's delta \cite{bartlett2006hedging,hagan2020bartlett} 
in this case ($\beta=1$).
A similar approximation also exists for the one-factor Rough Bergomi model 
(which includes the Brownian Bergomi model as a special case).

While our hedging problem is posed with discrete trading times and transaction costs, 
these exact or approximate explicit analytical strategies should be close 
to optimal for frequent trading without transaction costs in the parameter domains 
where the approximation performs very well. Otherwise, it should certainly 
serve as a useful reference point.

\section{Decision processes}

\subsection{Settings}

Many problems such as the ones considered here can be cast in the context of an agent interacting 
with some environment based on some decision policy that 
tries to optimize (maximize or minimize, as the case may be) some function of a sum of step-wise, 
initial, and final contributions ("rewards").\footnote{Instead 
of optimizing a function of the sum, one can also optimize over some characteristics of the 
distribution of the sum, such as a CVaR.}

To specify abstractly such settings, one needs to specify the state of the environment, 
the set of possible inputs to the agent, and the set of possible actions
by the agent. 
One also needs to specify how the environment transitions to a new state 
with new observations in reaction to an action (and this transition 
can be random and parts of the state might not be affected by the action) 
and what contribution each action and step will make to the sum of rewards (and 
that contribution might be stochastic as well) \cite{powell2022reinforcement}.

In some situations, the state and the evolution of the environment might 
not be completely captured. This makes such a setting very hard to handle and 
standard results do not apply and we will not discuss such settings further. 
The transitions and rewards might potentially depend on all the states and
actions that came before, presenting Nonmarkovian problems. Most algorithms 
and theory are presented for the Markovian case where both transitions and
rewards only depend on the most recent previous state and the current action, 
and we will only discuss this setting here. In at least some cases it is 
possible to exactly or approximately capture path dependency by extending 
the state space (such as for up or down Barrier options keeping track of 
running maximum or minimum or whether the corresponding barrier has been 
breached yet or not) and then using Markovian methods. There are circumstances
under which the agent only observes some part or function of the state etc., 
called partially observable decision processes. While such settings might 
be needed to treat settings in which there are latent parts of the state 
(such as an unobserved stochastic volatility), we will not cover their setting 
or theory here.

One sometimes introduces a reward discount factor $\gamma \in (0,1]$ if 
optimizing earlier rewards or costs is more important than optimizing later 
contributions, with $\gamma=1$ corresponding to the undiscounted case. 
(One could potentially allow $\gamma \geq 1$ to put more emphasis on later
contributions.) Such $\gamma$ also might help with the convergence or 
numerical behavior of the sum but a $\gamma$ not equal to 1 does change the 
problem and its solution in general. In our setting, we are mostly interested 
in the $\gamma=1$ case and might look at values close to 1 for numerical
purposes. 
 
Markov decision processes (MDPs) with deterministic rewards can be specified by the following:
\begin{itemize}
\item \textit{state space} $\mathcal{S}$: describes the state of the evolution 
of the environment and what the agent takes as input to make a decision
\item \textit{action space} $\mathcal{A}$:  describes the set of agent actions
\item \textit{Transitions}:
\begin{itemize} 
\item \textit{Transition probability} $\mathbb{P}:\mathcal{S}\times\mathcal{A}\times\mathcal{S}\longrightarrow[0,1]$ or \textit{transition density} $p:\mathcal{S}\times\mathcal{A}\times\mathcal{S}\longrightarrow \mathbb{R}^+$:   
$\mathbb{P}(s, a,s')$ denotes the probability of the state transitioning 
from $s$ to $s'$ as a result of taking the action $a$, and the 
next state $s'$ would be chosen based on the probability $\mathbb{P}(s, a,\cdot)$, or
\item \textit{Transition function} with explicit randomness: 
$s'=f(s,a,\xi)$ with $\xi$ describing 
the randomness and stochastic impact that impacts the transition between $s$ 
and $s'$.
\end{itemize}
\item Deterministic \textit{rewards} $r:\mathcal{S}\times\mathcal{A}\times\mathcal{S}\longrightarrow\mathbb{R}$: reward
assigned to the transition $(s, a,s')$.  
\item $\gamma$: the \textit{reward discount factor}.
\end{itemize}

The set of trajectories or realizations of MDP can be 
described as follows. One first starts with an initial 
state $s_0$ that could be fixed or given as a random 
variate or otherwise varying. Then, starting with time
index $t=0$, $s_t$ is given, agent chooses an action $a_t$, 
and environment generates a new state $s_{t+1}=f(s_t,a_t,\xi_t)$ 
and a reward $R_{t+1}=r(s_t,a_t,s_{t+1})$. 
Since there is randomness at 
least in the transition to the new state, $s_{t+1}$ 
and $R_{t+1}$ will be (at least partially) random and 
their distributions will depend on  the actions taken. 
The agent's decisions are generated from a policy $\pi$. 
A deterministic policy is a function $\pi:\mathcal{S}\longrightarrow\mathcal{A}$ 
and a stochastic policy assigns a distribution $\pi(.|s_t)$ 
over the set of actions where actions are drawn from $a_t\sim\pi(s_t)$.

$\mathbb{E}_{\pi}[.]=\mathbb{E}_{a_t=\pi(s_t)}[.]$ respectively
$\mathbb{E}_{\pi}[.]=\mathbb{E}_{a_t\sim\pi(.|s_t)}[.]$ denote the expectation under the randomness 
of the state transition if the actions $a_t$ of the agent are chosen according to policy $\pi$
for deterministic or stochastic policies, respectively. 

The \textit{return} of a MDP starting at time index $t$ is defined as
\begin{equation*}
G_t = \sum_{k=0}^{\infty}\gamma^{t+k}R_{t+k+1},
\end{equation*}
The return satisfies the recursion, $G_t = R_{t+1}+\gamma^{t+1}G_{t+1}$.

\subsection{Value functions and action-value functions}

For a fixed policy $\pi$, the \textit{value function} 
assigns to each state the expected return, conditional
on starting from state and selecting actions according 
to policy $\pi$ thereafter,
\begin{equation}
V^{\pi}(s) = \mathbb{E}_{\pi}\left[G_t|s_t=s\right].    
\end{equation}

Markov property gives a fundamental recursive structure 
to the value function, the Bellman 
equation for the value function:
\begin{align}
V^{\pi}(s) = R_{t+1}+\gamma\sum_{s'}\mathbb{P}^{\pi}(s,s')V^{\pi}(s'), \label{bellman_eq}
\end{align}
where, $\mathbb{P}^{\pi}(s,s')=\mathbb{P}\left(S_{t+1}=s'|s_t=s, a_t\sim\pi(s_t)\right)$.

The return of an MDP is a random variable, so one can take (conditional) expectations of its
second moment:
\begin{equation}\label{return_2nd_moment}
M^{\pi}(s) \coloneqq \mathbb{E}_{\pi}\left[G_t^2|s_t=s\right].
\end{equation}

It is possible to formulate a Bellman equation for the expectation of the second moment 
of the return   (derived in \cite{sobel1982variance} for a discounted infinite horizon with finite states 
and for the finite horizon in \cite{tamar2016learning}) as follows: one has the expansion
\begin{align*}
G_t^2 &= (R_{t+1}+\gamma G_{t+1})^2\\
            &= R^2_{t+1}+2\gamma R_{t+1}G_{t+1}+\gamma^2 G_{t+1}^2.
\end{align*} 

Therefore, one can re-write the expectation of the second moment of the return as,

\begin{align*}
M^{\pi}(s) &= \mathbb{E}_{\pi}\left[G_t^2|s_t=s\right]= 
\mathbb{E}_{\pi}\left[R^2_{t+1}+2\gamma R_{t+1}G_{t+1}+\gamma^2 G_{t+1}^2\right]\\
           &= \mathbb{E}_{\pi}\left[R^{(M)}_{t+1}\right]+ 
             \mathbb{E}_{\pi}\left[\gamma^2 M^{\pi}(s_{t+1})\right],
\end{align*} 
where,
\begin{equation*}
R^{(M)}_{t+1} = R^2_{t+1}+2\gamma R_{t+1}G_{t+1}.
\end{equation*}
By defining,
\begin{equation*}
r_{\pi}^{(M)}(s) \coloneqq\mathbb{E}_{\pi}\left[R_{t+1}^{(M)}|s_t=s\right] 
\end{equation*}
the Bellman equation for the expectation of the second moment takes the form,
\begin{equation}\label{bellman_eq_2nd_moment}
M^{\pi}(s)= r_{\pi}^{(M)}(s)+\sum_{s'}\mathbb{P}^{\pi}(s,s')\gamma^2 M(s').
\end{equation}

We introduce another concept, closely related to the value function.
The action-value function or the \textit{Q-function} at state $s$ and 
action $a$ is the value of taking the action $a$ and following the policy $\pi$ after that, 
\begin{equation}
Q^{\pi}(s,a) = \mathbb{E}\left[\sum_{t=0}^T\gamma^t R_{t+1}\middle|s_0=s, a_0=a\right].
\end{equation}
The value function can be expressed as the expectation of the 
Q-function across all possible actions,
\begin{equation}
V^{\pi}(s) = \mathbb{E}\left[Q^{\pi}(s,a)\middle|a_t\sim\pi(.|s)\right].
\end{equation}

The Q-function satisfies the Bellman expectation equation,
\begin{equation}\label{Q_Bellman_equation}
Q^{\pi}(s,a) = \mathbb{E}\left[ R_{t+1} + \gamma Q^{\pi}(s', a') \;\middle|\; s_t = s, a_t = a\right].
\end{equation}

Standard reinforcement learning tries to find an 
optimal policy that maximizes the expected return,
$\pi^{\ast,std}$ which would give the maximum 
action-value function $Q^{\ast}_(s,a)$ for any $s$ and $a$,
\begin{equation}
Q^{\ast}(s,a) = Q^{\pi^{\ast,std}}(s,a) \ge Q^{\pi}(s,a).
\end{equation}
There is a version of Bellman equation for the $Q^{\ast}$-function:
\begin{equation}
Q^{\ast}(s,a) = \mathbb{E}\left[ R_{t+1} + \gamma \max_{a'} Q^{\ast}(s', a') \;\middle|\; s_t = s, a_t = a\right]. \label{bellman_optimality_equation}
\end{equation}

Similar to (\ref{return_2nd_moment}), one can define 
an action-value function for the expectation of the second moment as:
\begin{equation}
K^{\pi}(s,a) \coloneqq  \mathbb{E}_{\pi}\left[G_t^2|s_t, a_t\right]
\end{equation}
and write that out as,
\begin{equation}\label{second_moment_q}
K^{\pi}(s,a) = \mathbb{E}\left[ R_{t+1}^2 + 2\gamma R_{t+1}Q^{\pi}(s', a') + \gamma^2 K^{\pi}(s', a') \;\middle|\; s_t = s, a_t = a\right].
\end{equation}
As expected, the expectation of the second moment depends on that of the first moment. 
Therefore, to achieve a good approximation of the expectation of the second moment, 
a good estimate of the expectation of the first moment, i.e. the Q-function, is required. 
In section \ref{ddpg_second} we will use (\ref{second_moment_q}) to find a 
policy that optimizes the second moment of Q-function. 
Similar ideas have been proposed by \cite{cao2021deep} to 
optimize a combination of the return of the MDP and its variance.

\subsection{Final quadratic hedging as Markov decision problem}

One way to fit quadratic hedging into an MDP/reinforcement-type learning set-up is 
to find such definitions of $\gamma$ and $R_t$ respective $R(s,a,s')$ so that 
$G_0 = \sum_{t=0}^N \gamma^t R_{t+1} = \tilde{L}_T$ or $G_0 = L_T$ 
(for deterministic interest rates, those two have the same optimizers). 
One typically uses $\gamma=1$ or very close to 1 since $\gamma$ cannot be independently chosen. 
One could try to learn strategies for smaller $\gamma$ (or larger $\gamma$) 
and then move $\gamma$ closer to 1. Some reinforcement algorithms only work 
(or can be proven to work) for $\gamma<1$ and certain assumptions on rewards.
We would then need a setting that satisfies those requirements. 
The methods we implemented here seem to work well enough for $\gamma$ equal or very close 
to 1. 
Then, minimizing $G_0^2$ will correspond to minimizing $\tilde{L}_T^2$.

We thus set $\gamma=1$, for $i=1,\cdots,N-1$ (with $T=t_N$)
\begin{equation}
R_i= H^S_{t_i} \left(\tilde{S}_{t_{i+1}}-\tilde{S}_{t_i}\right) - \mathsf{TC}(H^S_{t_i},H^S_{t_{i-1}},\tilde{S}_{t_i})
\label{cashflowform}
\end{equation}
and  
\begin{equation}
R_N=-\mathsf{TC}(0,H^S_T,\tilde{S}_{T}) - P_T(\tilde{S}_T).
\end{equation}

It is also possible to introduce some reference value function or 
process $\mathsf{Ref}$ with $\mathsf{Ref}_T=P_T(\tilde{S}_T)$ and 
use it to "shift" rewards around:
\begin{equation}
R_i= H^S_{t_i} \left(\tilde{S}_{t_{i+1}}-\tilde{S}_{t_i}\right) - \mathsf{TC}(H^S_{t_i},H^S_{t_{i-1}},\tilde{S}_{t_i}) + (\mathsf{Ref}_{t_i} - \mathsf{Ref}_{t_{i-1}}) \label{accountingform}
\end{equation}
where  $\mathsf{Ref}_{t_{-1}} = 0$. 
$\mathsf{Ref}_t$ will telescope out and lead to the same $G_0$ and $G_0^2$ 
(and same optimizers) but the formulation with $\mathsf{Ref}$ 
might be more amenable to particular reinforcement learning approaches. 
We did not apply this idea here since we were able to run the implemented reinforcement learning 
approach without introducing such a function or process 
which also means that one does not need to compute or define it. 

In \cite{cao2021deep}, (\ref{cashflowform}) is called the cash flow formulation, 
while (\ref{accountingform}) is called the accounting formulation. 

\subsection{Methods}

Standard reinforcement learning uses Bellman equations, together with temporal difference learning, 
and/or policy gradient lemma, in various types of methods \cite{fathi2023comparison}.
There are model based approaches in which the specification of the transitions and rewards 
are used and the sums or integrals are computed exactly or with some approximation so 
that all (or, at least, many) transitions and rewards are taken into account at once. 
However, this is only possible in some cases, often requires a finite set of states, 
and will often taken considerable computing power and memory. 
There are also various model free approaches that work based on 
observed transitions and rewards (single or a few observed steps each) 
rather than entire trajectories. Reinforcement learning algorithms 
try to learn value and action-value functions and a policy (or using 
a greedy or $\epsilon$-greedy policy derived from the action-value functions). 
For standard reinforcement learning, these methods are discussed at many 
places, we only point to \cite{fathi2023comparison} and work cited therein
as well as \cite{powell2022reinforcement} for a treatment and discussion
of reinforcement learning approaches and stochastic optimization approaches
for a variety of settings. However, there are many introductions to reinforcement 
learning available.  
We will discuss reinforcement-type learning approaches that optimize 
the expectation of the second moment in the next section. 

There are methods that work based on empirical/MC 
full-trajectory simulation of the sum of rewards, 
parametrizing the agent's policy as deep neural networks, 
and optimizing the policy by taking gradient or stochastic gradient 
steps (or extensions thereof, such as ADAM). We call these methods 
deep trajectory based stochastic optimal control (DTSoC) methods. 
In these methods, one does not (need to) learn value or action-value functions. 
We will discuss such methods in the section after the next. 

If the evolution of the environment as a system of SDEs 
and the controlled SDE can be captured in forward-backward stochastic 
differential/difference equations (FBSDE) and the objective function 
has appropriate initial, step-wise, and/or final terms, 
one can apply deep learning methods for FBSDE, which need limited 
model information and/or outputs but are otherwise generic. 
We will discuss such methods in a later subsection.

\section{Reinforcement Learning}

We refer to \cite{fathi2023comparison} for a description 
of Q-learning, policy gradient methods, and deep 
deterministic policy gradient methods, to maximize 
the return of the MDP with standard reinforcement learning. 
One can implement a variant of Q-learning that learns both 
$Q$ and $K$ functions. Here, we will present, implement,
and use a variant of  deep policy gradients that learns a 
$Q$ function, a $K$ function, and a 
policy that minimizes the variance, i.e., minimizes the $K$ function.

\subsection{Deep Deterministic Policy Gradients Variant for K-function}\label{ddpg_second}
In this section, we will propose an algorithm which instead of optimizing
$Q$-function optimizes $K^{\pi}(s,a)$, 
the variance of the return rather than the expected return. 
Our proposed algorithm involves the same ideas as DDPG to approximate the $Q$-function. 
However, there are two major differences between our algorithm and the original DDPG algorithm: 
1) The original algorithm includes one network to approximate the $Q$-function for the policy and one network to 
 approximate the policy and it updates the parameters of the $Q$-network by minimizing 
 the mean squared error of the difference between its approximation and target values.  
 However, our algorithm includes an extra network $K_\omega (a, s)$ with 
 set of parameters $\omega$ to estimate $K^{\pi}(s,a)$. 
 We update the parameters $\omega$ by minimizing the mean squared error between the output of the 
 $K_\omega$ and its target values. 
 The target values for the K-network are different from target values of $Q$-network 
 and are generated using (\ref{second_moment_q}).  
2) The policy in the original DDPG is updated toward choosing the actions that 
 optimizes the Q-function of the policy at a specific state $s$, however, in our algorithm 
 we update the policy to choose the action that optimizes (minimizes) the K-function - the 
 expectation of the second moment of the return. The details of our method are provided in Algorithm \ref{main_alg}.
See \cite{cao2021deep} for a similar method that minimizes a combination of $Q$ (mean) and $K$ 
(variance) functions.
 
\begin{algorithm}
\caption{Deep Deterministic Policy Gradient Algorithm for Second Moment of Return}\label{main_alg}
\begin{algorithmic}
\State Initialize policy parameters $\theta$, Q-function parameters $\phi$, K-function parameters $\omega$
\State Set target parameters equal to main parameters $\theta_{\text{targ}}\leftarrow\theta$, $\omega_{\text{targ}}\leftarrow\omega$, $\phi_{\text{targ}}\leftarrow\phi$
\State Initialize replay buffer $\mathcal{M}$
\For{$\text{episode}=1,\cdots,K$}
\State Initialize a random process $\mathcal{N}$ for exploration
\State Observe initial state $s_1$ 
\For{$t=1,\cdots,T(=\text{time horizon})$}
\State Select action $a_t=\mu_{\theta}(s)+\mathcal{N}_t$
\State Execute action, receive reward $r_t$ and state transition $s_{t+1}$
\State Append the transition to the replay buffer $\mathcal{M}$
\State Sample a mini-batch of transitions $B= \{(s_i, a_i, r_i,s_{i+1})\}$ from the replay buffer $\mathcal{M}$
\For{each transition tuple and  $s'=s_{i+1}$}
\State $\text{target}_{Q} = r+\gamma Q_{\phi_{\text{targ}}}\left(s',\mu_{\theta_{\text{targ}}}(s')\right)$ 
\State $\text{target}_{K} = r^2+\gamma^2K_{\omega_{\text{targ}}}\left(s',\mu_{\theta_{\text{targ}}}(s')\right) + 2\gamma rQ_{\phi_{\text{targ}}}\left(s',\mu_{\theta_{\text{targ}}}(s')\right)$
\EndFor
\State \textbf{Q-update:} Get updated $Q$-function network weights $\phi^{\text{updated}}$ by one-step gradient descent,
\begin{equation*}
\nabla_{\phi}\frac{1}{|B|}\sum_{(s, a, r,s')\in B}\left(Q_{\phi}(s,a)-\text{target}_{Q}\right)^2
\end{equation*}
\State \textbf{K-update:} Get updated $K$-function network weights $\omega^{\text{updated}}$ by one-step gradient descent,
\begin{equation*}
\nabla_{\omega}\frac{1}{|B|}\sum_{(s, a, r,s')\in B}\left(K_{\omega}(s,a)-\text{target}_{K}\right)^2
\end{equation*}
\State \textbf{Policy update:} Get updated policy network weights $\theta^{\text{updated}}$ by one-step gradient descent,
\begin{equation*}
\nabla_{\theta}\frac{1}{|B|}\sum_{s\in B}K_{\omega}(s,\mu_{\theta}(s))
\end{equation*}
\State Update target networks
\begin{equation*}
\phi_{\text{targ}}\leftarrow\rho\phi_{\text{targ}}+(1-\rho)\phi^{\text{updated}}
\end{equation*}
\begin{equation*}
\theta_{\text{targ}}\leftarrow\rho\theta_{\text{targ}}+(1-\rho)\theta^{\text{updated}}
\end{equation*}
\begin{equation*}
\omega_{\text{targ}}\leftarrow\rho\omega_{\text{targ}}+(1-\rho)\omega^{\text{updated}}
\end{equation*}
\EndFor
\EndFor
\end{algorithmic}
\end{algorithm}

\section{Deep Trajectory-Based Stochastic Optimal Control}

Deep Trajectory-Based Stochastic Optimal Control (DTSOC), proposed in \cite{han2016deepsc} 
(also see \cite{Deep_stochastic} for an exposition), is a method for solving stochastic control problems 
through formulating the control problem as optimizing over a computational graph, with the sought controls represented as (deep) neural networks. 
The approximation power of deep neural networks can mitigate the curse of dimensionality for solving dynamic programming problems.

We briefly review the setup of the method here. Consider a stochastic control problem given by the following underlying stochastic dynamics,
\begin{equation}
s_{t+1} = f(s_t,a_t,\xi_t)
\end{equation}
where, $s_t$ is the state, $a_t$ is the control (agent's action) and $\xi_t$ is a stochastic disturbance impacting the period between time index $t$ and $t+1$.  

In the models for the hedging instruments derived from (discretized) SDEs, 
the $\xi_t$ will be the Brownian increments in the (discretized) SDEs for 
the time step from index $t$ to $t+1$, $\Delta W_t$ as a discretization of $dW_t$. 
The price of the hedging instrument as well as the amount of them held would be part of the 
state and the action would either directly give the new amount to be held or 
an increment or rate that would allow the new amount to be computed. 

We assume that the actions are given as deterministic or stochastic feedback controls 
$a_t = \pi_t(s_t|\theta_t)$
or
$a_t \sim \pi_t(\cdot|s_t,\theta_t)$.
One can extend the state $s_t$ with path-dependent extra state that can be computed from current and previous state, action, and disturbances; 
and also with particular precomputed features that might lead to more efficient training of agents or more efficient agents which 
also extends the set of controls that can be written as feedback controls. 

The actions can be constrained to come from a set of admissible functions: 
\begin{equation*}
a_t\in\mathcal{A}_t=\left\{a_t: 
g(s_t,a_t)=0, h(s_t,a_t)\geq0\right\},
\end{equation*} 
where $h(s_t,a_t)$ and $g(s_t,a_t)$ are inequality and equality constraints. We assume that these constraints are already taken into account in the feedback controls such 
that $\pi_t$ will be an admissible action or give a distribution over admissible actions. 

We assume step-wise contributions given by $c_t=c_t(s_t,a_t,s_{t+1})$ and also a final contribution $c_T(s_T)$.
 
Given a deterministic or probabilistic policy in feedback form that gives admissible actions, we generate 
an episode 
\[
s_0, a_0, s_1, a_2, \cdots, a_{T-1}, s_T
\]
and obtain a total contribution
\begin{equation}
C= \sum_{t=0}^{T-1} c_t(s_t,a_t,s_{t+1}) + c_T(s_T).
\end{equation}

The stochastic optimal control problem now minimizes (or maximizes) 
a loss function $l$ of the expected total contribution, conditional on starting state $s_0$. If $s_0$ is not 
fixed, this will be a function of $s_0$. 
We thus try to minimize $\E[l(C)|s_0=s]$ or $\E[l(C)]$ varying the policies $\pi_t$. 
For quadratic hedging, the loss function is $l(x)=x^2$. Other loss functions can be considered as long as they can be meaningfully optimized over 
by appropriate mini-batch approaches. 

With some given functional form (such as DNN) with an appropriate parametrization (for example,
determine weights and biases while activation functions are fixed for complete feed-forward DNN) as deterministic policy, we obtain
\begin{equation}
 \E\left[l(C)\right] = \E\left[ l\left(\sum_{t=0}^{T-1} c_t\left(s_t,\pi_t(s_t|\theta_t),s_{t+1}\right) + c_T(s_T)\right) \right] =: {\cal L}\left(\left\{ \theta_t\right\}_{t=0}^{T-1} \right) =: {\cal L}(\Theta) \label{DNN_loss}
\end{equation}

One now jointly optimizes over all policies $\{\pi_t\}^{T-1}_{t=0}$ respective over all parameters of such $\left\{ \theta_t\right\}_{t=0}^{T-1}$ to optimize the loss 
function as applied to the total contribution (if $s_0$ is not fixed, this will also be a function of $s_0$ and we would need to take an appropriate expectation over $s_0$ or keep $s_0$ as a parameter).

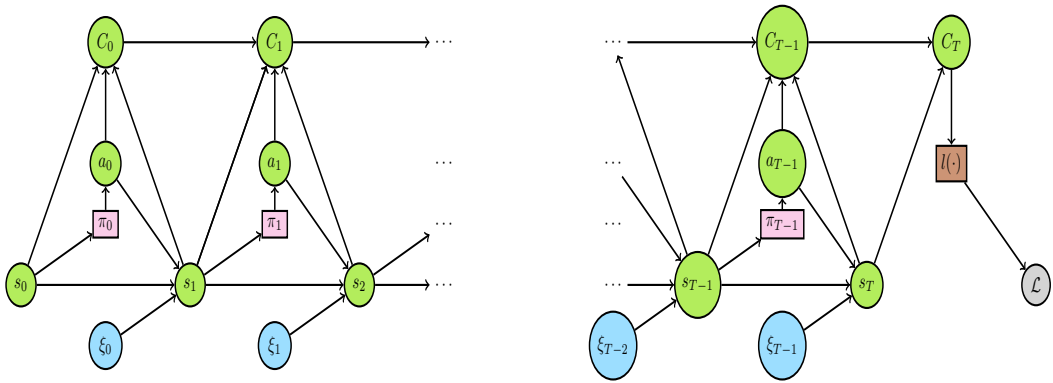
\begin{figure}[h!]
\begin{center}
\resizebox{140mm}{50mm}{
\begin{tikzpicture}

\node[draw,circle,fill=inchworm] (s0) at (0,2) {$s_0$};
\node[draw,circle,fill=columbiablue] (xi0) at (2,1) {$\xi_0$};
\node[draw,fill=classicrose] (pi0) at (2,3) {$\pi_0$};
\node[draw,circle,fill=inchworm] (a0) at (2,4) {$a_0$};
\node[draw,circle,fill=inchworm] (c0) at (2,6) {$C_0$};
\node[draw,circle,fill=inchworm] (s1) at (4,2) {$s_1$};

\draw[->] (s0) to (pi0);
\draw[->] (s0) to (c0);
\draw[->] (pi0) to (a0);
\draw[->] (a0) to (c0);
\draw[->] (a0) to (s1);
\draw[->] (xi0) to (s1);
\draw[->] (s0) to (s1);
\draw[->] (s1) to (c0);

\node[draw,circle,fill=columbiablue] (xi1) at (6,1) {$\xi_1$};
\node[draw,fill=classicrose] (pi1) at (6,3) {$\pi_1$};
\node[draw,circle,fill=inchworm] (a1) at (6,4) {$a_1$};
\node[draw,circle,fill=inchworm] (c1) at (6,6) {$C_1$};
\node[draw,circle,fill=inchworm] (s2) at (8,2) {$s_2$};

\draw[->] (c0) to (c1);
\draw[->] (s1) to (c1);
\draw[->] (s1) to (pi1);
\draw[->] (s1) to (c1);
\draw[->] (pi1) to (a1);
\draw[->] (a1) to (c1);
\draw[->] (a1) to (s2);
\draw[->] (xi1) to (s2);
\draw[->] (s1) to (s2);
\draw[->] (s2) to (c1);

\node (somes) at (10,2) {$\mathbf{\cdots}$};
\node (somepi) at (10,3) {$\mathbf{\cdots}$};
\node (somea) at (10,4) {$\mathbf{\cdots}$};
\node (somec) at (10,6) {$\mathbf{\cdots}$};

\draw[->] (s2) to (somes);
\draw[->] (s2) to (somepi);
\draw[->] (c1) to (somec);

\node (somes1) at (14,2) {$\mathbf{\cdots}$};
\node (somepi1) at (14,3) {$\mathbf{\cdots}$};
\node (somea1) at (14,4) {$\mathbf{\cdots}$};
\node (somec1) at (14,6) {$\mathbf{\cdots}$};

\node[draw,circle,fill=columbiablue] (xitm2) at (14,1) {$\xi_{T-2}$};

\node[draw,circle,fill=inchworm] (stm1) at (16,2) {$s_{T-1}$};

\draw[->] (xitm2) to (stm1);
\draw[->] (somes1) to (stm1);
\draw[->] (somea1) to (stm1);
\draw[->] (stm1) to (somec1);

\node[draw,circle,fill=columbiablue] (xitm1) at (18,1) {$\xi_{T-1}$};
\node[draw,fill=classicrose] (pitm1) at (18,3) {$\pi_{T-1}$};
\node[draw,circle,fill=inchworm] (atm1) at (18,4) {$a_{T-1}$};
\node[draw,circle,fill=inchworm] (ctm1) at (18,6) {$C_{T-1}$};
\node[draw,circle,fill=inchworm] (st) at (20,2) {$s_T$};

\draw[->] (somec1) to (ctm1);
\draw[->] (stm1) to (pitm1);
\draw[->] (stm1) to (ctm1);
\draw[->] (pitm1) to (atm1);
\draw[->] (atm1) to (ctm1);
\draw[->] (atm1) to (st);
\draw[->] (xitm1) to (st);
\draw[->] (stm1) to (st);
\draw[->] (st) to (ctm1);

\node[draw,circle,fill=inchworm] (ct) at (22,6) {$C_T$};

\draw[->] (ctm1) to (ct);
\draw[->] (st) to (ct);

\node[draw,fill=antiquebrass] (losscomp) at (22,4) {$l(\cdot)$};

\draw[->] (ct) to (losscomp);

\node[draw,circle,fill=lightgray] (loss) at (24,2) {$\cal L$};

\draw[->] (losscomp) to (loss);

\end{tikzpicture}
}  
\end{center}
\caption{Computational Graph for DTSOC. \label{DTSOC_CG}}
\end{figure}

The controls at each time step could be stacked into a computational graph with a loss function given in \eqref{DNN_loss}. 
For each roll out of the control problem, this computational graph takes the sequence of disturbances $\{\xi_t\}_{t=0}^{T-1}$ as 
input and gives the loss function as applied to the total contribution inside the expectation in \eqref{DNN_loss} as output. 
Figure \ref{DTSOC_CG} shows the computational graph to compute the loss ${\cal L}$ and has the following features:

\begin{itemize}
\item The deterministic policy at time step $t$ is represented by some network with appropriate architecture (shown in a pink box) $s_t\rightarrow a_t$ with parameters $\theta_t$ that are trainable (can be optimized over)
\item The transition of the system to a new state, $(s_t, a_t)\rightarrow s_{t+1}$ based on the system dynamics is encoded in the connections from $s_t$, 
$a_t$, and the random disturbance $\xi_t$ (shown in blue) to $s_{t+1}$.
\item $s_t$ and $s_{t+1}$ will be input to $c_t$ as shown in the graph
\item Defining the cumulative contribution up to time $t$ as,
\begin{equation*}
C_t = \sum_{\tau=0}^t c_\tau(s_\tau,a_\tau,s_{\tau+1}),
\end{equation*}
the horizontal connections on top of the network, $(s_t,a_t,C_t)\rightarrow C_{t+1}$ sums up the time $t$ contribution and gives the total accumulated contribution $C_T$
at the end of the episode (when $t=T$).
\item The total accumulated contribution $C_T$ is then passed through a loss function (shown in light brown) and gives the loss $\cal L$ (shown in gray) as final result.
\end{itemize}

Note that based on a discretization $\{0=t_0<t_1,\cdots,t_p=T\}$ of the time horizon, 
the computational graph will have $p$ layers (with $p$ embedded DNN) and $p\times\sum_{t=0}^{T-1}N_t$ trainable parameters. 
After the loss ${\cal L}$ has been computed as in the above computational graph, 
standard deep learning frameworks such as TensorFlow or PyTorch can now use the 
computational graph to generate path-wise gradients with respect to all trainable parameters.

Pseudo-code for DTSOC is given in Algorithm \ref{DSOC_training} below. 
\begin{algorithm}[H]
\caption{Training Procedure for DTSOC}\label{DSOC_training}
\begin{algorithmic}
\State \textbf{Initialize:} Network weights $\{\theta_m, m=0,\cdots,T-1\}$
\While {epoch$\leq EPOCH$}
\State $nbatch=0$
\State $batchloss=0$
\While {$nbatch\leq \text{batch size}$}
\State $m=0$
\State $C_{-1}=0$
\While {$m<T$}
\State $a_m = \pi^{\theta_{m}}(s_{m})$\
\State $s_{m+1} = f (s_{m}, a_{m},\xi_m)$ with sampled noise $\xi_m$. 
\State $C_m=C_{m-1} + c_m(s_{m},a_{m},s_{m+1})$
\State $\text{m++}$
\EndWhile
\State $C_T=C_{T-1} + c_T(s_T)$
\State $batchloss+=l(C_T)$
\EndWhile
\State Calculate loss for batch $$\text{Loss}=batchloss/(\text{batch size})$$
\State Calculate gradient of $\text{Loss}$ with respect to $\theta$
\State Back propagate updates for $\{\theta_m,m=0,\cdots,T-1\}$ 
\State $\text{epoch++}$
\EndWhile
\State \textbf{return} optimized weights $\{\theta^*_m,m=0,\cdots,T-1\}$ 
\end{algorithmic}
\end{algorithm} 

The contributions $c_t(s_t,a_t,s_{t+1})$ correspond to the step-wise rewards in the reinforcement learning 
approaches. Just like there, only the sum counts in terms of optimal solutions. 

The "deep hedging" paper \cite{buehler2018deep,buehler2019deep} presents a trajectory based empirical deep stochastic optimal control
approach to minimizing some global objectives related to replication and/or risk management of some final payoff. It mentions a 
version with loss function in equation (3.3) and mentions some multi-dimensional quadratic hedging example in section 5.4, but does not
present detailed discussion or tests for quadratic hedging.  
 
\subsection{Relationship to FBSDE Formulation of Stochastic Control}

We here repeat and adapt the discussion from \cite{fathi2023comparison}. In general, one can consider 
a stochastic control problem in which some functional defined by running and final costs 
(which depend on the evolution of some controlled forward SDE and on the control) is optimized over that control. 
This leads  to coupled \textit{forward backward stochastic differential equations} (FBSDE) and non-linear PDEs 
(See \cite{perkowski2011backward} or \cite{pham2009continuous} for an introductory treatment). In our setting, the control 
that the agent tries to optimize does not impact the forward SDEs describing the evolution of the prices of the 
instruments, it only impacts the trading strategy, leading to a controlled backward SDE only.  

With $X_t$ being the factors and prices for the (hedging) instruments and $Y_t$ being the value of the hedging strategy, 
we have the system (see \cite{hientzsch2019introduction,hientzsch2021intro})\footnote{
In this subsection, we use notation from the FBSDE literature as adapted to the pricing and hedging domain and do not follow the generic 
notation for RL or trajectory-based approaches. The state $s_t$ in RL or trajectory-based approaches 
would contain $X_t$, $Y_t$, $\Pi_t$, $J_t$, and whatever is needed to compute terms and costs (or equivalent information), the action/control
would be some parametrization of $\Pi_t$, the stochastic disturbance $\xi_t$ would be the $dW_t$ or $\Delta W^i$.
}, 
\begin{equation}
dX_{t} = \mu\left( t,X_{t} \right)dt + \sigma\left(t,X_{t}\right)dW_{t}
\end{equation}
\begin{equation}
 dY_{t} = - f\left( t,X_{t},Y_{t},\Pi_{t} \right)dt\  + \Pi_{t}^{T}
\sigma\left(t,X_{t}\right) dW_{t}
\end{equation}
where $\Pi_{t}$ plays the role of a control or strategy and the functional to be optimized (typically, minimized) is 
\begin{equation}
J^F(\Pi_t,\Pi^{\mathsf{final}},\ldots) = E\left(\int_0^T
{\mathsf{rc}}(s,{X}_s,Y_s,\Pi_s) ds + {\mathsf{fc}}({X}_T,Y_T,\Pi^{\mathsf{final}})\right).
\end{equation}
For the example of an European option and quadratic hedging, 
$Y_T$ has to replicate the appropriate payoff $g(X_T)$ in the mean square sense, i.e. $\mathsf{fc}({X}_T,Y_T,\cdot) = \left(Y_T-g(X_T)\right)^2$.
If the market is complete, $Y_T$ will be perfectly replicable and thus the exact loss function or final cost does not 
matter as long as it is zero when $Y_T$ is perfectly replicated.  

One can define 
\[ J_t =  \int_0^t {\mathsf{rc}}(s,{X}_s,Y_s,\Pi_s) ds \]
or 
\[ dJ_t = {\mathsf{rc}}(t,{X}_t,Y_t,\Pi_t) dt \]
and add it to the stochastic system, looking for a minimum of 
\[ \E \left( J_T +  {\mathsf{fc}}({X}_T,Y_T,\Pi^{\mathsf{final}})\right).\]

One can derive FBSDE characterizing the optimal controls (both primal and dual/adjoint) as well as
PDEs characterizing them, but we will here concentrate on approaches that directly optimize 
over the given system for $X_t$, $Y_t$, $\Pi_t$, and $J_t$. 

Upon time-discretization, one obtains stochastic control problems defined on (controlled) FBS$\Delta$E ($\Delta$ standing for "difference") where now the 
running cost can depend on the forward and backward components and the control at both the beginning and end of each time-period. 

Applying a simple Euler-Maruyama discretization for both $X_t$ and $Y_t$, we
obtain 
\begin{equation}
X_{t_{i+1}} = X_{t_i} + \mu(t_i,X_{t_i}) \Delta t_i + \sigma^T(t_i,X_{t_i})
\Delta W^i
\end{equation}
\begin{equation}
Y_{t_{i+1}} = Y_{t_i} - f\left( t_i,X_{t_i},Y_{t_i},\Pi_{t_i} \right) \Delta 
t_i + \Pi^T_{t_i} \sigma^T(t_i,X_{t_i}) \Delta W^i
\end{equation}
This can be used to time-step both $X_t$ and $Y_t$ forward. 

Now, quadratic hedging means that one minimizes the squared differences $E(|Y_T-g(X_T)|^2)$ and the form of this final loss function 
matters since one in general can no longer perfectly replicate $g(X_T)$.

Similarly, the running costs need to be accumulated
\begin{equation}
J_{t_{i+1}} = J_{t_i} + {\mathsf{rc}}(t_i,{X}_{t_i},Y_{t_i},\Pi_{t_i}) \Delta t_i
\end{equation}
and the stochastic optimal control problem will try to minimize
\[ \E \left( J_T +  {\mathsf{fc}}({X}_T,Y_T,\Pi^{\mathsf{final}})\right).\]

A time-discrete setting allows one to incorporate more general transaction costs for $Y_t$ \cite[Section 7.2]{hientzsch2019introduction} 
by more complicated generators $f$
\begin{equation}
Y_{t_{i+1}} = Y_{t_i} - f_{\Delta t} \left( t_i,\Delta t_i,
X_{t_i},X_{t_{i+1}},Y_{t_i},Y_{t_{i+1}},\Pi_{t_i},\Pi_{t_{i+1}} \right) 
+ \Pi^T_{t_i} \sigma^T(t_i,X_{t_i}) \Delta W^i
\end{equation} 
and also more complicated running costs 
\begin{equation}
J_{t_{i+1}} = J_{t_i} + {\mathsf{rc}}\left(t_i,\Delta t_i,{X}_{t_i},X_{t_{i+1}},Y_{t_i},Y_{t_{i+1}},\Pi_{t_i},\Pi_{t_{i+1}}\right) \Delta t_i,
\end{equation}
which could include running costs that depend on the profit and loss of some strategy across the corresponding time interval.

In \cite{weinan2017deep,hientzsch2019introduction,hientzsch2021intro,ganesan2019pricingbarriers,ganesan2022pricingbarriers,liang2019deep,liang2021deep}, 
path-wise deepBSDE methods for such problems are discussed, at least applied to pricing and risk management where 
there is only a final cost (or a cost at the earlier of reaching a barrier or maturity) - as in the quadratic hedging setup considered here.

DeepBSDE methods represent the strategy $\Pi_t$ as a DNN depending on appropriate state $X_t$ (or features computable from such state). 
Path-wise forward deepBSDE methods generate trajectories of $X$ and $Y$ starting from initial values $X_0$ and $Y_0$ 
according to the current strategy. They then use stochastic gradient  descent type approaches such as ADAM to improve the strategy
until an approximate optimum is reached. 
If the initial wealth $Y_0$ is not given, it will be determined by the optimization as well. 
If the starting value $X_0$ of the risk factor vector is fixed, $Y_0$ would be a single value, 
otherwise it would be a function of $X_0$. The optimization problem would typically represent
this function as a DNN. 

Derived so far for certain kinds of final costs or where one attempts to replicate the final payoff as well as possible, path-wise backward deepBSDE 
methods make the same assumptions, but on each generated forward trajectory of $X$, they start with an appropriate final value of $Y_T$ ($Y_T=g(X_T)$ for
the final payoff case), compute a corresponding trajectory of $Y_t$ by stepping backward in time, and try to minimize the range of $Y_0$. 
It appears that the strategies computed by the backward deepBSDE methods also perform well when used and tested with the quadratic hedging 
loss function in a forward stepping approach. 

Quadratic hedging for European options has been considered with
 forward and backward path-wise methods in \cite{liang2019deep,liang2021deep}
for linear pricing and in \cite{yuhientzsch2019backward,yuhientzsch2023backward} 
for nonlinear pricing (differential rates) while forward path-wise methods were
introduced earlier by  \cite{weinan2017deep}. 
The barrier option case is treated with forward methods in 
\cite{ganesan2019pricingbarriers,ganesan2022pricingbarriers}. 

One difference between the setup discussed in this section and 
other approaches considered in the paper is that here the backward SDE or  S$\Delta$E 
is written in such a way that it uses parts of the forward model (i.e. $ \sigma^T(t_i,X_{t_i}) \Delta W^i$)
and might not as written satisfy self-financing exactly but only up to discretization accuracy. In this way, this is a (more) model-based approach. 
However, one can rewrite the BSDE for $Y$ so that it only uses the stochastic increment of $X$ (i.e., written in $dX$ rather than using model details about $X$) 
and one can rewrite the BS$\Delta$E so that it only requires observations of $X$ at trading times $t_i$ and perfectly preserves self-financing 
similarly to what we wrote in earlier subsections, obtaining methods that will be more similar to the ones discussed there.

\section{Experimental Setup and Model Specification}\label{experimental_setup}
We consider the example of a European call option with strike price $K$ and maturity $T$ on a non-dividend-paying stock. 
The strike price and option maturity are considered as fixed parameters. 
It is assumed that the risk-free rate is zero and that the option position is held until maturity. 
Rebalancing of the hedging portfolio is allowed at fixed (often regular) times 
and trades are subject to transaction cost proportional to the trade size. 
The trained hedging agent is expected to learn to hedge an option with this specific set of parameters. 
It is possible to train parametric agents that can hedge a parametrized set of options (such as calls with various strikes), but we will not do so here. 
We assume the Black-Scholes or SABR model for the simulation environment where the stock dynamics is given by
\begin{equation}
dS_{t} = \mu S_t \; dt + \sigma_t S_t \; dW_t \label{SDE:Stock},
\end{equation}
with either a constant $\sigma_t=\sigma$ (Black-Scholes) or a stochastic volatility 
$\sigma_t$ given by the SDE
\begin{equation}
d\sigma_t = \frac{\eta}{2} \sigma_t dW^2_t
\end{equation} 
and trading (buying and selling) of stock incurs a transaction cost which is assumed to have the functional form,
\begin{equation}
\text{cost}(S_t, \delta H_t) = \alpha | S_t\delta H_t|,
\end{equation}
where $\delta H_t$ is the change in the stock position.

The default parameters for the stock, the option and transaction costs are as given in Table \ref{BS-defaultparameters}
with the additional parameters for the SABR model as in Table \ref{SABR-defaultparameters}.
\begin{table}[htb]
\begin{center}
\begin{tabular}{| l || l | } \hline
Parameter & Value \\ \hline
$\mu$ & 5 \% (rate of return) \\ \hline
$\sigma$ & 20 \% (Black-Scholes volatility) \\ \hline
$ir$ & 0.0 \% (interest rate) \\ \hline
$S_0$ & 100 \\ \hline
$K$ & 100 \\ \hline
$T$ & 30 (option maturity- days) \\ \hline
$\alpha$ & 0.001 (transaction cost parameter)\\ \hline
\end{tabular}
\caption{Default parameters of stock, option, and transaction cost.\label{BS-defaultparameters}}
\end{center}
\end{table}

\begin{table}[htb]
\begin{center}
\begin{tabular}{| l || l | } \hline
Parameter & Value \\ \hline
$\sigma_0$ & 20 \% (initial volatility) \\ \hline
$\eta$ & 0.95 \% (volatility of volatility) \\ \hline
$\rho$ & 0.5 \% (correlation of stock and volatility Brownian) \\ \hline
\end{tabular}
\caption{Additional default parameters for SABR model.\label{SABR-defaultparameters}}
\end{center}
\end{table}

\subsection{State Space Selection and Model Architecture}\label{model_architecture}

There are at least two parts of state - one part describes the state of the environment and its evolution;
the other part is provided as input to the training and trained agent; and these two 
parts certainly overlap. Certain features or transformations of features can be added 
that might enabled faster training or more optimal agents. One can also train agents that 
are robust against certain changes in parameters, but we will not do so here. 

The state of the environment and the action are characterized by the following variables, 
given at each time step:
 
\begin{itemize}
\item Time $t$, $0\leq t\leq T$,
\item Stock price at time $t$,
\item Current stock holding\footnote{Given by or impacted by agent's action in previous step(s).}  $H_t$.
\end{itemize}

One can parametrize the trading strategy directly by specifying $H_t$, the quantity of stock held at time $t$
or by assuming that amount rebalanced is proportional to time passed and parametrizing the rebalancing rate. 
The quantity of stock held should be Markovian and thus directly 
parametrizing it might have certain advantages.  
The two parametrizations are related through the equation, 
\begin{equation}\label{rate_to_ratio}
H_{t_{m+1}}^{\theta_{m+1}}=H_{t_{m}}^{\theta_{m}}+ \dot{H}_{t_m}^{\theta_{m}}\Delta_t.
\end{equation}
We choose to parametrize the quantity of stock held.
This is similar to how other implementations in the literature have used relatively shallow deep networks with three layers for parametrizing 
each individual control (see \cite{han2016deepsc}).
RELU was used as the nonlinear activation function in all hidden layers. 

Deep-QH agent is a feed-forward neural network with three hidden layers with 10, 15 and 10 neurons,
with batch-normalization in each layer. The final output transformation is chosen as linear. 

For RL-QH, the critic networks to learn the $Q$ and $K$ function and 
the actor network all have two hidden layers with 32 and 64 neurons,
with batch-normalization before and after each layer. 
The $Q$ and $K$ function networks have linear output transformation 
while the actor/policy network has a sigmoid output transformation.

\subsection{Model Training}\label{model_training}
The model was trained on a computer with 8 core i7- 11850 CPU $@$ 2.50GHz processor and 32.0 GB memory.

For the RL-QH model, both actor and critic are trained with respect to a mean squared error (MSE) loss 
with the TensorFlow implementation of ADAM optimizer, actor learning rate and critic 
learning rate are set to $\expnumber{1}{-4}$ and $\expnumber{1}{-4}$ respectively. 
The smoothing parameter for updating the weights of the target network is $\expnumber{1}{-5}$. 
The RL-QH model was trained over 50,000 episodes.  
The above parameters were kept fixed for both zero and non-zero transaction cost regimes. 

The deep-QH model is trained using ADAM optimization with Pytorch off-the-shelf parameters. 
Batch normalization and dropout ($p=0.25$) were applied to each layer. 
The learning rate was fixed at $\expnumber{1}{-3}$ initially and was decreased dynamically as training epochs progressed. 
The model was trained for 50,000 episodes.

\section{Results for Black-Scholes}

\begin{figure}[H]
\begin{center}
\fbox{\includegraphics[width=0.6\textwidth,keepaspectratio]{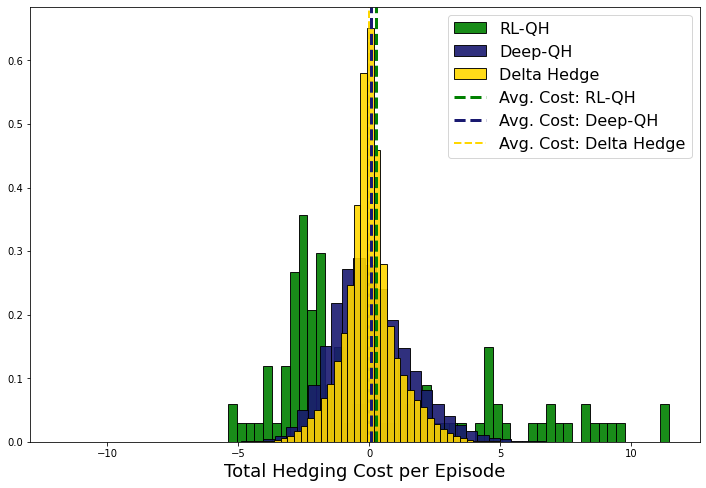}}
\caption{Histogram of hedging cost at maturity, zero transaction cost. The $x$-axis is the total hedging cost during the life of the option. Positive values denote loss and negative values denote profit.\label{benchmark_RL_vs_deepQH_no_TC}}
\end{center}
\end{figure}

We trained the Deep-QH and RL-QH agents as described above and compared them against the Delta
hedge agent (which is variance optimal in the zero transaction cost case),
for an option with 30 days to maturity. We show histograms
for the final mismatch between the final value of the trading strategy versus the required final 
payoff in figure \ref{benchmark_RL_vs_deepQH_no_TC} . 
Negative values (to the left) mean that the strategy was worth more than the payoff and
leading to a profit after paying out the required payoff. Positive values (to the right) mean
that the strategy was worth less than the payoff and would lead to a loss after paying out 
the required payoff. We observe that all three agents have very similar average profit or loss,
with Delta hedge being a bit more concentrated at zero profit or loss, with the Deep-QH spread 
out somewhat more. RL-QH sometimes leads to bigger profits but also allows larger losses, controlling 
P\&L not as tightly. 

\subsection{Longer Maturity}

\begin{figure}[H]
\centering
\fbox{\includegraphics[width=0.8\textwidth,keepaspectratio]{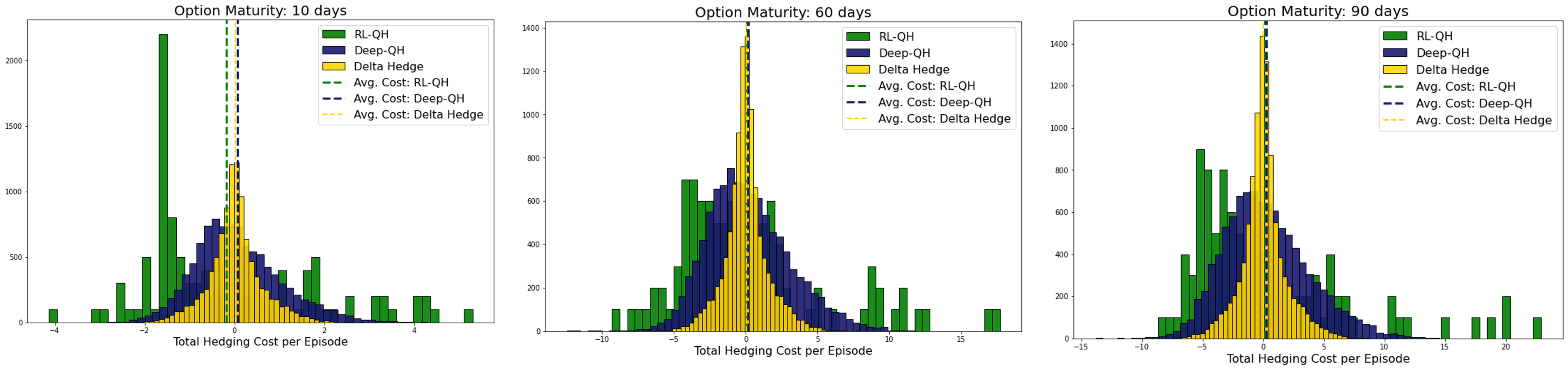}}
\caption{Histogram of hedging cost for RL-QH, deep-QH and Delta hedging strategies with increasing option maturity. The $x$-axis is the total hedging cost during the life of the option. Positive values denote loss and negative values denote profit.
\label{benchmark_RL_vs_deepQH_increasing_maturity}}
\end{figure}

Here, we trained the agents to hedge options with 
shorter maturity (10 days) or longer maturities (60 days and 90 days)
and show the results in figure \ref{benchmark_RL_vs_deepQH_increasing_maturity}. 
For shorter maturities, RL-QH outperforms on average while giving very 
similar results to 30 days for longer maturities. RL-QH consistently shows 
a far wider spread, showing that it does not control tails as well
as the Deep-QH agent or the Delta hedge agent. All hedging agents 
allow a larger spread or variance for longer option maturities and 
as before, the Deep-QH agent allows a somewhat wider spread but still 
substantially more controlled than the RL-QH agent. 

\subsection{Increasing Volatility}

\begin{figure}[H]
\centering
\fbox{\includegraphics[width=0.8\textwidth,keepaspectratio]{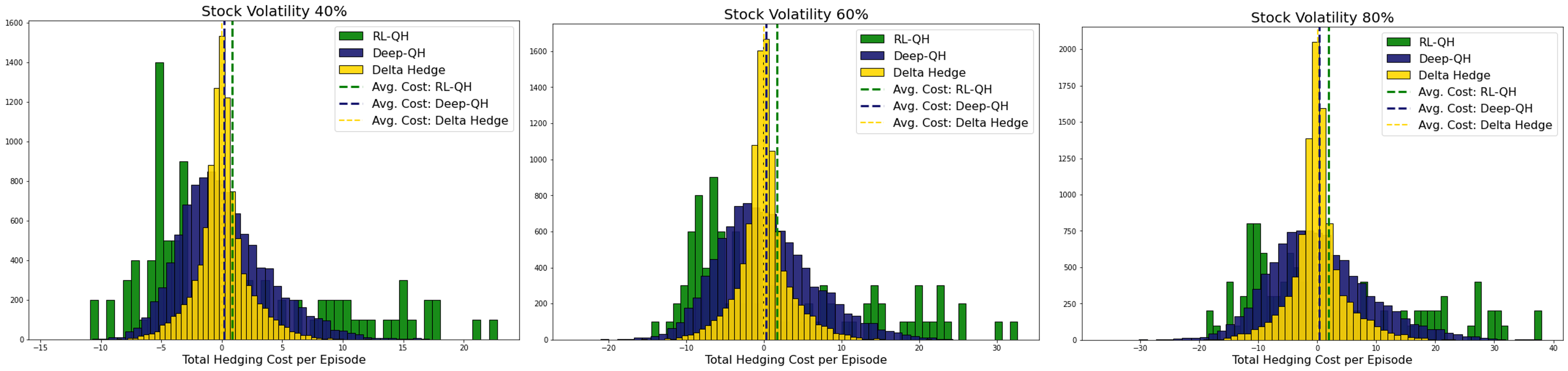}}
\caption{Histogram of hedging cost for RL-QH, deep-QH and Delta hedging strategies with increasing Volatility. The $x$-axis is the total hedging cost during the life of the option. Positive values denote loss and negative values denote profit.}
\label{benchmark_RL_vs_deepQH_increasing_volatility}
\end{figure}

Here, we compared settings with increasing volatility of the underlying stock 
while keeping option maturity at 30 days, showing the results in figure
\ref{benchmark_RL_vs_deepQH_increasing_volatility}.
With increasing volatility, the RL-QH agent does worse on average, realizing 
fewer larger profits while still allowing larger losses. 
The variance for all agents increases with increasing volatility. The relative 
performance and shape relationship between Deep-QH and Delta hedge remains the 
same. 

\subsection{Increasing Transaction cost}

\begin{figure}[H]
\centering
\fbox{\includegraphics[width=0.8\textwidth,keepaspectratio]{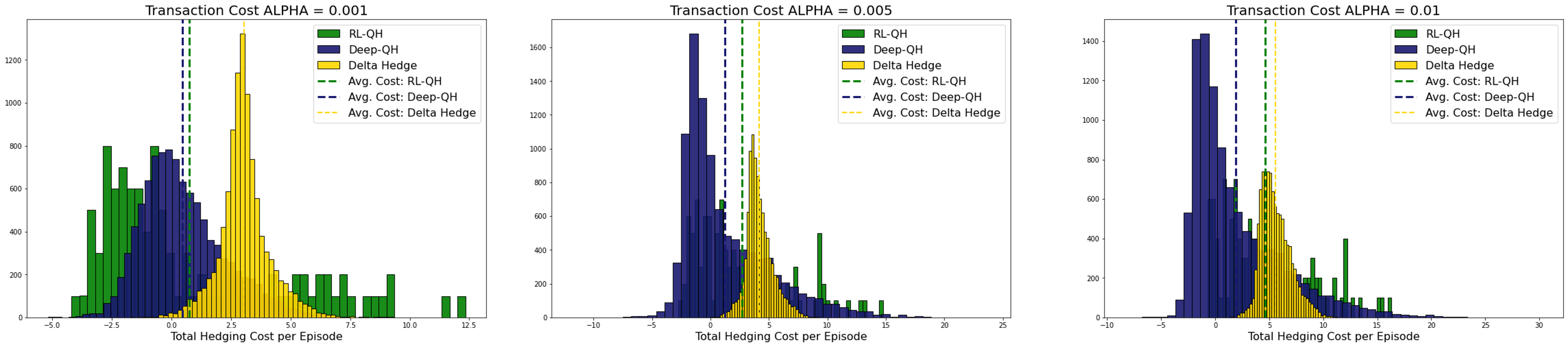}}
\caption{Histogram of hedging for RL-QH, deep-QH and Delta hedging strategies with increasing transaction cost. The $x$-axis is the total hedging cost during the life of the option. Positive values denote loss and negative values denote profit.}
\label{benchmark_RL_vs_deepQH_increasing_TC}
\end{figure}

Figure \ref{benchmark_RL_vs_deepQH_increasing_TC} shows the 
performance of the RL-QH, Deep-QH and Delta hedge agents
with increasing transaction costs. 
One can observe that both RL-QH and Deep-QH systematically
outperform the Delta hedge agent, with the Deep-QH agent performing
increasingly better against the RL-QH agent. 
While both RL-QH and Deep-QH have a peak corresponding to at 
least moderate gains and a substantial percentage of trajectories
that end in gains, there is a heavier tail of larger losses,
compare to the tails of the Delta hedge agent, for large enough
transaction costs. While the behavior of the Deep-QH seems to
be relatively smooth, RL-QH allows a set of more uncontrolled losses.

\section{Results for SABR}

\begin{figure}[H]
\begin{center}
\fbox{\includegraphics[width=0.6\textwidth,keepaspectratio]{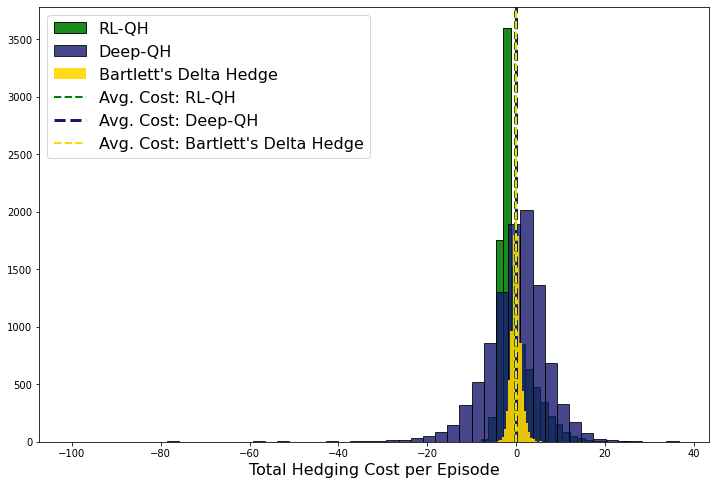}}
\caption{Histogram of hedging cost at maturity under SABR model. The $x$-axis is the total hedging cost during the life of the option. Positive values denote loss and negative values denote profit}
\label{benchmark_RL_vs_deepQH_SABR}
\end{center}
\end{figure}

Here, the RL-QH and Deep-QH agents were trained on trajectories 
generated by the log-normal SABR  
model, where the stochastic volatility is not observed by the agents.
We also implemented an agent that implements Bartlett's Delta
hedge with the approximation for implied volatility as described 
in an earlier section.  
The results are shown in figure \ref{benchmark_RL_vs_deepQH_SABR}.
One can observe that after training, RL-QH and Deep-QH agents 
on average perform as well as the approximately variance optimal 
Barlett's Delta hedge agent. Both RL-QH and Deep-QH allow larger
variance than the Barlett's Delta hedge agent. Given that 
stochastic volatility is unobserved by the agents, at least 
part of this variance could be explained by that. We intend
to closer investigate this in future work. 

\subsection{Model Robustness and Generalization}

\begin{figure}[H]
\centering
\fbox{\includegraphics[width=0.6\textwidth,keepaspectratio]{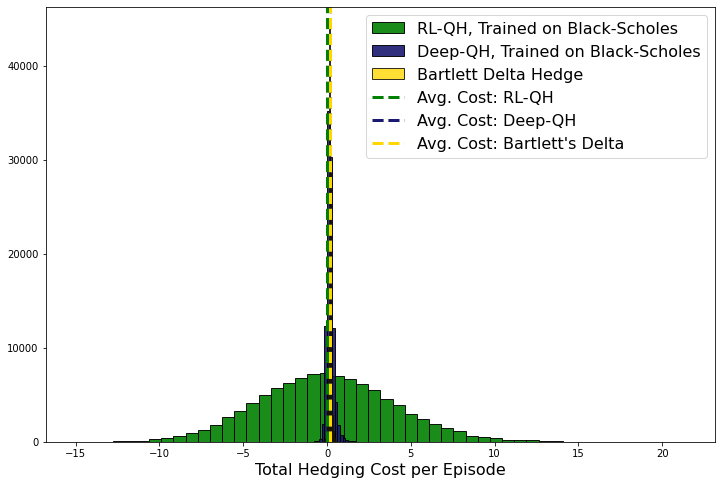}}
\caption{Histogram of hedging cost at maturity, for RL-QH, deep-QH and Bartlett's Delta hedging 
strategies under the SABR dynamics. The RL-QH and deep-QH agents are trained under the 
Black-Scholes environment. The $x$-axis is the total hedging cost during the life of 
the option. Positive values denote loss and negative values denote profit.}
\label{benchmark_RL_vs_deepQH_robustness}
\end{figure}

As a simple model robustness test that tests whether agents trained on simpler 
models perform sensibly on more complicated models, we trained RL-QH and
deep-QH agents on Black-Scholes models corresponding to the initial volatility
in the log-normal SABR model and tested them against the variance-optimal 
Bartlett's Delta hedge agent when trajectories are generated by the log-normal
SABR model. Histograms\footnote{Figure 
\ref{benchmark_RL_vs_deepQH_SABR} shows results for a larger number 
of option contracts while figure \ref{benchmark_RL_vs_deepQH_robustness} shows
results for one option contract, thus the ranges and variances of the 
RL-QH and deep-QH agents trained on Black-Scholes are substantially larger than 
for the agents trained on SABR as in figure \ref{benchmark_RL_vs_deepQH_SABR} -
the range and variance of the Bartlett's Delta hedge agent in both figures
are the same.} for the final profit or loss are shown 
in figure \ref{benchmark_RL_vs_deepQH_robustness}.
One can observe that all three agents lead to very similar average performance. 
Deep-QH agent's performance is spread out more than the Barlett's Delta performance,
but still tightly peaked close to zero loss or gain. While RL-QH
has similar average performance, its performance shows a very wide 
variance. It would be interesting to investigate this further to see how
this performance comparison depends on parameters and also whether 
training on Black-Scholes models with varying parameters will lead 
to improved performance on the SABR test case.

\section{Conclusion and Future Work}

We implemented and studied deep trajectory-based stochastic optimal control and deep reinforcement approaches 
to minimize the variance of the final hedging P\&L. In particular, we implemented and applied an extension 
of the DDPG Actor-Critic approach to the case where the expectation of the second moment of the return 
is optimized rather than the expected return. We trained and tested on the Black-Scholes model and 
the log-normal SABR model. Without transaction costs, variance optimal strategies are known - for Black-Scholes,
Delta hedging is optimal, while Bartlett's Delta hedging with the exact implied volatility is variance optimal 
in the log-normal SABR model when only the underlier is used for hedging. With a good approximation for the 
implied volatility as available for the SABR model, Barlett's Delta hedging with that approximation is very close
to variance optimal. Deep-QH and RL-QH agents match the (approximate) variance optimal strategies in average 
cost with comparable (deep-QH) or wider (RL-QH) variance and range, with RL-QH allowing a wider range of both 
extreme gains and losses. For non-zero and increasing transaction costs, both RL-QH and deep-QH outperform 
the variance optimal Delta hedging on average, with deep-QH doing so more consistently than RL-QH in 
the Black-Scholes case.  Similar results are seen for log-normal SABR which is and example for an incomplete 
market (if only hedged with stock). We finally tested agents trained on Black-Scholes model with the 
initial volatility from the SABR model but tested them on SABR trajectories. On average, RL-QH and deep-QH
still match the performance of Bartlett's Delta hedge, however showing larger variance (for deep-QH)
and dramatically larger variance (RL-QH). 

This paper and these results suggest areas of additional work. A more complete study of the SABR
model and other more complicated and incomplete models, studying further behavior with increasing transaction 
costs and with varying maturity, strike, and volatility would be in order. It would be interesting to see whether other 
architectures, algorithmic choices in the RL algorithm, and hyper-parameters and choices in the training 
would allow better control of variance and outliers for the RL approaches and improve the deep-QH results 
even further. A second set of questions concerns latent factors such as the stochastic volatility in SABR. 
What would the impact be if the stochastic volatility would be treated as observable and provided to the
agents? What would the impact be of adding additional hedge instruments such as a variance swap or one
or several European options to complete the market? Would a more robust training (over varying volatility 
parameters, for instance) under the Black-Scholes model lead to better performing agents under the SABR model
with better variance control? Would adding volatility estimates that are computed from the observed
or generated sequence of spot prices as input features to the agent allow that training of agents that
are more robust against model choice and specification? 

Also, a set-up that allows easier exploration of models, trading strategies, and objective functions, 
such as an extension of a generic simulation framework as used in \cite{polala2023parametric}
would make such studies easier. 

Finally, how would a transaction-cost-aware delta strategy (as the Leland model and strategy for Black-Scholes)
perform compared to RL and deep trajectory-based stochastic optimal control approaches? Would training 
of RL and deep-QH approaches on different transaction costs and giving transaction cost parameters as 
input to those agents lead to improved training and behavior in the presence of transaction costs?

\section{Acknowledgments}

The author acknowledges and appreciates the contributions by 
Ali Fathi (who during these contributions was working  at Wells Fargo) who implemented the reinforcement 
learning and deep trajectory-based stochastic optimal control algorithms,
the Black-Scholes and log-normal SABR settings, and generated the results shown in this paper. 
The author appreciates the assistance of Abdolghani Ebhrahimi of Wells Fargo
with the specification, adaptation, and running of the reinforcement learning algorithms for the second 
moment of the MDP return. 

\newpage
\bibliographystyle{alpha}
\bibliography{rlanddsoc}

\end{document}